\documentclass{emulateapj}

\usepackage{natbib}
\def\HII{{\ion{H}{2}}}
\def\HI{{\ion{H}{1}}}
\def\OII{[{\ion{O}{2}}]}

\def\OIIIHb{[{\ion{O}{3}}]/H$\beta$}
\def\OIII5007Hb{[{\ion{O}{3}}]~$\lambda5007$/H$\beta$}
\def\4959_5007{[\ion{O}{3}]~$\lambda \lambda$4959,5007}
\def\OIII49595007{[\ion{O}{3}]~$\lambda \lambda 4959,5007$}
\def\ratioR23{([\ion{O}{2}]~$\lambda$3727 +
[\ion{O}{3}]~$\lambda\lambda$4959,5007)/H$\beta$}
\def\R23{${\rm R}_{23}$}
\def\dS23{${\rm S}_{23}$}

\def\Msun{${\rm M}_{\odot}$}

\def\NII{[{\ion{N}{2}}]}
\def\OIIIOII{[\ion{O}{3}]/[\ion{O}{2}]}

\def\NIIOII{[\ion{N}{2}]/[\ion{O}{2}]}

\def\OH{$\log({\rm O/H})+12$}
\def\NIISII{[\ion{N}{2}]/[\ion{S}{2}]}
\def\NIIOII{[\ion{N}{2}]/[\ion{O}{2}]}

\def\ratioS23{([\ion{S}{2}]~$\lambda \lambda$6717,31 +
[\ion{S}{3}]~$\lambda\lambda$9069,9532)/H$\beta$}
\def\NIIHa{[\ion{N}{2}]/H$\alpha$}
\def\SIIHa{[\ion{S}{2}]/H$\alpha$}
\def\OIHa{[\ion{O}{1}]/H$\alpha$}

\def\SII{[{\ion{S}{2}}]}

\def\Hb{{H$\beta$}}
\def\O4363{[{\ion{O}{3}}]~$\lambda$4363}
\def\OIII{[{\ion{O}{3}}]}

\def\Ha{{H$\alpha$}}

\def\L60{L$_{60}$}

\def\MB{${\rm M_{B}}$}

\shorttitle{}
\shortauthors{}

\begin{document}

\title{Metallicity and Nuclear Star Formation in Nearby Galaxy Pairs: Evidence for Tidally Induced 
Gas Flows}

\author{Lisa J. Kewley\altaffilmark{1}}
\affil {University of Hawaii}
\affil{2680 Woodlawn Drive, Honolulu, HI 96822}
\email {kewley@ifa.hawaii.edu}
\altaffiltext{1}{Hubble Fellow}

\author{Margaret J. Geller}
\affil{Smithsonian Astrophysical Observatory}
\affil{ 60 Garden Street MS-20, Cambridge, MA 02138}

\author{Elizabeth J. Barton}
\affil{University of California, Irvine}
\affil{Department of Physics and Astronomy,4154 Frederick Reines Hall,Irvine, CA 92697}


\begin{abstract}
We derive the first luminosity-metallicity relation for a large objectively selected sample of local galaxy pairs and we compare the pairs LZ relation with the relation for the 
Nearby Field Galaxy Survey (NFGS).   Galaxy pair members with small projected separations ($s<20$~kpc/h) have systematically lower metallicities ($\sim 0.2$~dex on average) than the field galaxies, or than more widely separated pairs at the same luminosity.  There is a strong correlation between metallicity and central burst strength in the galaxy pairs.   All five galaxies in the pairs sample 
with strong central bursts have close companions and metallicities lower than the comparable field galaxies.      Our results provide strong observational evidence for a merger scenario 
where galaxy interactions cause gas flows towards the central regions, carrying less enriched 
gas from the outskirts of the galaxy into the central regions.  The less enriched
gas dilutes the pre-existing nuclear gas to produce a lower metallicity than would be obtained prior to the interaction.   These gas flows trigger central bursts of star formation, causing strong central burst 
strengths, and possibly aiding the formation of blue bulges.  We show that the timescale and central gas dilution required by this scenario are consistent with predictions from hydrodynamic merger models.
\end{abstract}


\section{Introduction}
Galaxy interactions and mergers are fundamental to galaxy formation and evolution.  
The most widely supported merger scenario is based on the \citet{Toomre77} 
 sequence in which two galaxies lose their mutual orbital
energy and angular momentum to tidal features and/or an extended dark
halo and then coalesce into a single galaxy.  Tidal interactions and
associated shocks are thought to trigger star formation \citep[e.g.,][]{Bushouse87,Kennicutt87,Barnes04}. 
Theory predicts that, as a merger progresses, the galaxy disks become disrupted by tidal 
effects, causing gas flows towards the central regions where kpc-scale starbursts 
may be fueled \citep{Barnes96,Mihos96,Iono04,Springel05}.   Evidence for gas flows has been found in cold and 
warm gas in interacting galaxies \citep[e.g.,][]{Combes94,Hibbard96,Georgakakis00,Chen02,Marziani03,Rampazzo05}.  Large-scale gas flows combined with central bursts of star formation may be important for transforming merging galaxies into elliptical galaxies \citep[e.g.,][]{Kauffmann93,Kobayashi04,Springel05,Nagashima05}. 

The effect of galaxy interactions on star formation has been studied
for decades.  \citet{Morgan58} first discovered ``hotspots'' in
galaxies and \citet{Vorontsov59} published the
first catalogue of interacting galaxies.  In 1966, Arp published the famous Atlas of
Peculiar Galaxies, sparking much research into the properties of 
peculiar galaxies.   \citet{Sersic67} 
found that the galaxies harboring hotspots all have extreme blue colors 
in their centers, a property of the majority of the galaxies in both 
Vorontsov-Velyaminov and Arp's catalogues.  
The suggestion that tidal forces in interacting galaxies could trigger
bursts of star formation was first made by \citet{Larson78}, and since then,
numerous studies have provided evidence for interaction-induced
starburst activity \citep[see][ for a review]{Schweizer05}.   
Interacting galaxies have elevated UV emission \citep[][]{Petrosian78}, \Ha\ emission \citep[e.g.,][]{Balzano83,Kennicutt87,Bushouse87,Barton00}, IR emission \citep[e.g.,][]{Lonsdale84,Young86,Heckman86,Solomon88}, radio continuum  emission \citep[e.g.,][]{Hummel81,Condon82,Hummel90},
 and soft X-ray emission \citep[e.g.,][]{Read98}.  Mergers have been linked to the
extreme infrared luminosities seen in luminous infrared galaxies \citep[e.g.,][]{Soifer84,Armus87,Sanders96,Genzel98}, 
and the unusually blue colors observed in irregular galaxies \citep{Sersic67}. 

The observed relationship between star formation and galaxy interactions 
has been studied extensively, but we are just beginning to understand the metallicity
properties of merging galaxies.   The metallicity of normal disk galaxies is strongly correlated with 
galaxy mass.  The first mass-metallicity relation was found for irregular and blue compact galaxies \citep{Lequeux79,Kinman81}.  In subsequent work, luminosity was often used instead of mass because obtaining reliable mass estimates was difficult.   \citet{Rubin84} provided the first evidence that metallicity is correlated with luminosity in disk galaxies.  Further work using larger samples of nearby disk galaxies confirmed this result \citep{Bothun84,Wyse85, Skillman89,Vila92, Zaritsky94,Garnett02}.  The origin of the mass-metallicity relation for disk galaxies is unclear.   A mass-metallicity relation will arise if low mass galaxies have larger gas fractions than higher mass galaxies, as is observed in local galaxies \citep{McGaugh97,Bell00,Boselli01}.  Selective loss of heavy elements from galaxies in supernova-driven outflows can also account for the relation \citep{Garnett02,Tremonti04}.

In addition to providing information about previous episodes of star formation and mass-loss,  metallicity may provide clues into the dynamical effects of interactions on 
gas flows in galaxy pairs.   Isolated disk galaxies often 
display metallicity gradients \citep[see][for a review]{Shields90,Dinerstein96} which may be altered during a merger as a result of 
tidally-driven gas flows.  Metallicity gradients in interacting galaxies could potentially provide a unique method for finding interacting galaxies that have experienced strong tidal gas flows.
Few studies to date investigate the effect of galaxy mergers and interactions 
on the metallicity properties of galaxies.   Recent work in this area focuses on 
small numbers of galaxies, or on single galaxies.  
\citet{Donzelli00} analysed the optical spectra of 25 merging galaxies and find
that the spectra of mergers are more excited on average than interacting pairs.  Donzelli \& 
Pastoriza attributed this result to lower gas metallicity in the galaxy mergers, although it 
is not clear from their work what the cause of the lower gas metallicity could be.
\citet{Marquez02} used the \NIIHa\ ratio to study the metallicities in disk \HII\ regions in 13 non-disrupted
galaxies in pairs and a large sample of isolated galaxies.  They find that the \NIIHa\ ratios in the disk \HII\ regions of non-disrupted 
galaxy pairs are higher than for isolated galaxies,  but that there is no difference in nuclear \NIIHa\ 
ratios.   Recently, \citet{Baldi05} and \citet{Fabbiano04} 
investigated the Ne, Mg, Si, and Fe metallicities within the hot gas in 
X-ray luminous star forming regions of the Antennae galaxies.  They find that the metal ratios 
for these hot regions are consistent with enrichment of the gas by Type II supernovae.

These investigations provide the first clues into the impact of interactions on the 
metallicity gradients in galaxies.  A more comprehensive analysis of the metallicity properties 
in merging galaxies requires well-defined and carefully selected surveys.  
In this paper, we compare the metallicity and star formation 
properties of two samples that have been objectively selected from the CfA redshift catalog: (1)
the galaxy pairs sample of \citet{Barton03}, and (2) the Nearby Field Galaxy Survey \citep{Jansen00a}.
The objective sample selection enables powerful statistical comparisons to be made
between the two samples.  We describe and compare the two samples in Section~\ref{sample}.
In Section~\ref{class} we discuss our classification scheme for the removal of AGN and we compare 
the locus of the galaxy pairs with that of the NFGS on the standard optical 
diagnostic diagrams.   We investigate the ionization parameter differences between the 
pairs and the NFGS samples in Section~\ref{ionization}.
The metallicity estimates for the galaxy pairs and the NFGS
are in Section~\ref{metallicity}, which contains the first 
luminosity-metallicity (LZ) for galaxy pairs.
 We show that galaxy pairs have lower metallicities on average 
than field galaxies within the same luminosity range.  We investigate the relationship 
between metallicity and central burst strength and we 
examine whether metallicity is correlated with the presence of blue bulges for the galaxy pairs.   
In Section~\ref{discussion} we discuss our results in terms of the  merger 
scenario predicted by hydrodynamic simulations, and our conclusions are 
in Section~\ref{conclusions}.

Throughout this paper, we adopt the flat 
$\Lambda$-dominated cosmology as measured by the WMAP experiment 
($h=0.72$, $\Omega_{m}=0.29$; \citet{Spergel03}).

\section{Sample Selection} \label{sample}

We use two objectively selected local samples: the galaxy pairs of 
\citet{Barton00} and the Nearby Field Galaxy Survey \citep{Jansen00a}.  Ground-based
nuclear optical spectra are available for both surveys; the spectra include $\sim 10$\% of the total 
B-band light.    

\citet{Jansen00a} selected the NFGS from the CfA1 redshift survey 
\citep{Davis83,Huchra83}.  The galaxies were selected from 
 1 mag-wide bins of  $M_{Zwicky}$\footnote[1]{$M_{Zwicky}$ is approximately equal to $M_{B}$} and CfA1 morphological type to approximate the local galaxy 
luminosity function \citep[e.g.,][]{Marzke94}.  To avoid a strict diameter limit, which 
might introduce a bias against the 
inclusion of low surface brightness galaxies in the sample, Jansen et al. imposed a
radial velocity limit, ${\rm V_{LG} (km\,s^{-1}) > 10^{-0.19-0.2M_{Zwicky}}}$
(with respect to the Local Group standard of rest).  Galaxies in the
direction of the Virgo Cluster were excluded to avoid favoring a cluster population.  
The final 198-galaxy sample contains the full range in Hubble type and absolute 
magnitude present in the CfA1 galaxy survey and is a fair representation of the 
local galaxy population.   

\citet{Jansen00b} obtained nuclear spectra with a $3^{\prime \prime}$ slit centered on the 
nucleus and aligned along the major axis of each galaxy, sampling
$3''\times 6 \farcs84$.  The covering fraction of the nuclear spectra depends on 
the radial light profile and ranges between 0.4-72\% of the light within the 
$B_{26}$ isophote, with an average covering fraction of 10\%.  However for the sample
that we use in this study (described in Section~\ref{class}), the covering fraction ranges from 
0.4 -33\% with a mean covering fraction of 9\%.

We calibrated the nuclear fluxes to absolute fluxes 
by careful comparison with B-band surface photometry 
\citep[described in ][]{Kewley02b}.  
We corrected the \Ha\ and \Hb\ emission-line fluxes 
for underlying stellar absorption as described in \citet{Kewley02b}.

We used two methods to correct the NFGS emission-line fluxes 
for Galactic extinction, based on: 
(1) the HI maps of \citet{Burnstein84}, as listed in the Third Reference Catalogue of Bright Galaxies \citep{deVaucouleurs91}, 
and (2) the COBE and IRAS maps (plus the Leiden-Dwingeloo maps 
of HI emission) of \citet{Schlegel98}.  The average Galactic extinction is
E($B-V$)=$0.014 \pm 0.003$ (method 1) or E($B-V$)=$0.016 \pm 0.003$ (method 2).

\begin{figure}
\plotone{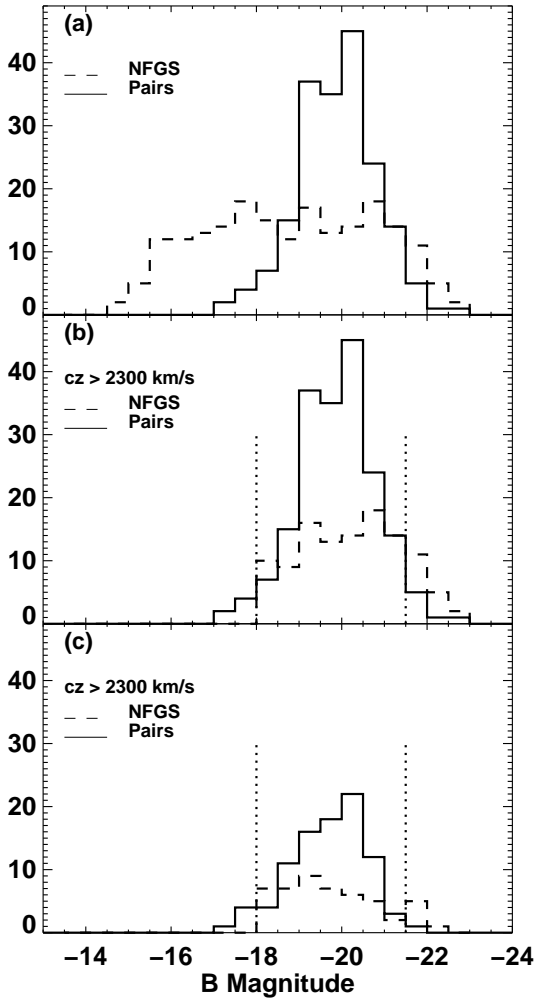}
\caption{The B-band luminosity distribution of (a) the galaxy pairs and the NFGS, 
(b) the galaxy pairs and the NFGS  with a recessional velocity limit $cz \geq 2300$~km/s, (c) the galaxy pairs and the NFGS that have measurable metallicities and are dominated by star-formation.
The NFGS contains a larger fraction of lower luminosity galaxies than the pairs sample.  For the purposes of metallicity comparisons, we compare galaxies within the same luminosity range of $-21.5 \leq {\rm M_{B}} \leq -18$.
\label{lum_hist}}
\end{figure}

\citet[][hereafter BGK00]{Barton00} selected a sample of 786 galaxy pairs and N-tuples from the CfA2 redshift survey, 
\citep{Geller89,Huchra90,Huchra95}.   The CfA2 redshift survey is magnitude-limited at $m_{Zw} = 15.5$.   These galaxies were selected with line-of-sight velocity 
separations $\Delta V \leq 1000$~km/s, projected separation $s \leq 50$~$h^{-1}$~kpc, and 
$cz \geq 2300$~km/s.  The lower velocity limit excludes the Virgo cluster and limits the angular sizes
of the galaxies.  
Of these 786 galaxies, BKG00 obtained nuclear spectra for 502 galaxies using a $3^{\prime \prime}$ slit centered on the nucleus of each galaxy.   The covering fraction of the spectra is approximately $ 10$\% of the galaxy 
light, comparable to the NFGS covering fraction.  The spectra were flux calibrated in a relative sense by BKG00 using standard stars.  The observations were not taken under photometric conditions, thus we have only relative calibration across each spectrum, suitable for line ratio diagnostics and equivalent width measurements.

Because the BGK00 sample selects only on projected separation and line of sight velocity separation in a complete magnitude limited sample,  the selection method is unlikely to produce correlations between pair separation and any spectroscopic or photometric property.  The BGK00 sample includes all pairs which satisfy the selection criteria, regardless of morphology or gas fraction.   

The 502-galaxy sample is primarily composed of pairs;  382/502 (74\%) galaxies are in pairs, 70/502 (14\%) galaxies are in groups of three, and 10/502 galaxies (10\%) are in groups of four or more.   A total of 190 galaxies in the pairs sample have B-band and R-band magnitudes measured by BKG00. 
 
Figure~\ref{lum_hist}a shows the B-band magnitude (${\rm M_{B}}$) distribution of the pairs and the NFGS.  The NFGS spans a much larger range of \MB, primarily because the NFGS sample was selected
to reflect the local galaxy luminosity function.  Because of this selection, low luminosity galaxies appear to be over-represented in the NFGS sample compared to the pairs.     The 
galaxy pairs were selected with  a lower recessional velocity limit of $cz \geq 2300$~km/s.  Figure~\ref{lum_hist}b gives the luminosity distribution after applying a lower $cz$ limit of $cz \geq 2300$~km/s to the NFGS.   Figure~\ref{lum_hist}b shows that for galaxies with $cz \geq 2300$~km/s, the NFGS has a luminosity distribution that is slightly shifted towards high luminosities.  This effect occurs because \citet{Jansen00a} applied a lower radial velocity limit that increases with luminosity for the NFGS.     For the purposes of metallicity comparisons, we compare galaxies within the same luminosity range of $-21.5 \leq {\rm M_{B}} \leq -18$ and with  $cz \geq 2300$~km/s. 
This selection ensures that (1) we compare similar size galaxies to minimise aperture effects 
\citep[see][for a discussion of aperture effects in the NFGS sample]{Kewley05a}, and (2) there are
sufficient numbers of galaxies in each 1-Mag luminosity bin for comparison.  Because galaxy metallicities correlate with global luminosity \citep[e.g.,][]{Kobulnicky99a,Melbourne02,Tremonti04,Kobulnicky04,Salzer05}, metallicities must be compared within the same luminosity range.  We use B-band  for our sample comparisons because both the pairs and the NFGS samples were selected from the magnitude limited CfA redshift catalogs based 
on the \citep{Zwicky61} catalog.  Therefore, the B-band is the consistent band to use for comparisons between samples.

\section{Classification} \label{class}

We use the nuclear optical emission-line ratios to classify the galaxies into \HII\ region-like 
or AGN-dominated classes.  
For galaxies with  \OIIIHb, \NIIHa, \SIIHa\ and (if available) 
\OIHa\ ratios, we used the theoretical optical classification scheme developed by 
\citet{Kewley01a}.  In this scheme, galaxies that lie above the theoretical ``extreme starburst 
line" in the standard optical diagnostic diagrams are AGN; those below the line are \HII\ region-like.  Galaxies in the AGN region 
in one diagnostic diagram but in the \HII\ region section of the other diagram are ``ambiguous" (AMB) galaxies.  A fraction of galaxies in the NFGS (35/198) and pairs sample 
(49/502) have \NIIHa\ ratios but immeasurable \OIII\ or \Hb\ fluxes in their nuclear spectra.  
We classify these galaxies as HII region-like if  log(\NIIHa)$<-0.3$, typical of starburst galaxies and \HII\ regions \citep[e.g.,][]{Kewley01b,Dopita00}.  We are unable to classify those galaxies with $-0.3\leq $log(\NIIHa)$\leq 0.0$ if \OIIIHb\ is not available because these line ratios can be produced by both AGN and starburst galaxies.  Galaxies without \OIIIHb\ but with strong log(\NIIHa)$>0.0$ are classed as AGN \citep[e.g., Figure~1 of ][]{Brinchmann04}.
The nuclear spectra of 124 NFGS galaxies contain sufficient line ratios for classification. 
Of these, the nuclear spectra of
105/124 (85\%) NFGS galaxies are dominated by star formation, 12/124 (10\%) 
are dominated by AGN, and 7/124 (5\%) are ``ambiguous''.   Because ambiguous galaxies are likely to contain both starburst and AGN  activity \citep[see e.g.,][]{Kewley01b,Hill99}, we do not include them in the following analysis.  
The remaining galaxies do not have sufficient emission-lines in their spectra to allow 
classification.   Many of these unclassified galaxies are ellipticals.  Out of the 105 star-forming
galaxies in the NFGS, a total of 43 galaxies have $cz \geq 2300$~km/s and  $-21.5 < {\rm M_{B}} < -18$.  These 43 galaxies comprise our comparison field galaxy sample.

A total of 212/502 of the pair members spectra contain emission lines suitable for classification.  
Of these, 153/212 (72\%) pair members are dominated by star formation, 
36/212 (17\%) are dominated by AGN, and 23/212 (11\%) have ambiguous line ratios.  Out of the 
153 star-forming galaxy pairs, a total of 92 galaxies have B-band photometry from \citet{Barton03}, and 86 galaxies have 
$-21.5 \leq {\rm M_{B}} \leq -18$.  These 86 galaxies comprise our "galaxy pairs" sample.  Although all of the galaxies within our pairs sample are members of a pair or n-tuple, our final 86-galaxy sample does not necessarily contain all members of each pair or n-tuple.  
We note that classification (and metallicity determination) with optical emission-lines selects against elliptical galaxies and galaxies with low gas fractions in the original 502-galaxy catalog of BKG00.
 
\begin{figure}
\plotone{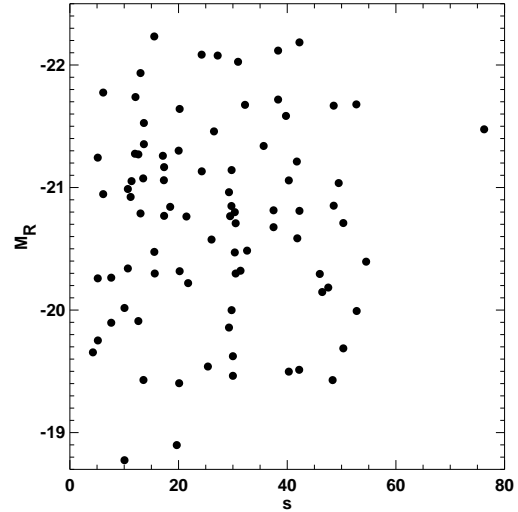}
\caption{Projected separation $s$ in units of kpc/h versus R-band luminosity for the entire pairs sample.  Filled circles denote the galaxy pairs that satisfy our emission-line selection criteria.
Clearly our selection does not bias the sample against massive galaxies at low projected separations.
\label{s_vs_RMag}}
\end{figure}

We check that our emission-line selected pairs sample has not biased the pairs against luminous high-mass (and high metallicity) systems as a function of projected separation in Figure~\ref{s_vs_RMag}.  Clearly there is no deficit of luminous pair members at low projected separations.

\begin{figure}
\plotone{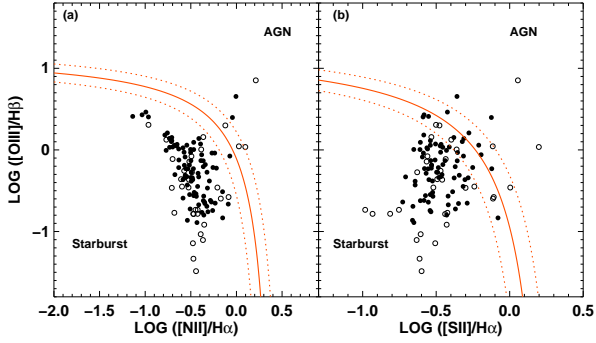}
\caption{The standard optical diagnostic diagrams showing the position of the nuclear line ratios of 
the galaxy pairs (filled circles) and the NFGS galaxies (unfilled circles) with $cz \geq 2300$~km/s and $-21.5 \leq {\rm M_{B}} \leq -18$.  The \citet{Kewley01a} 
theoretical classification line is shown in red.  The two dashed lines indicate the error range ($\sim 0.1$dex) of the classification line (solid line).   The star-forming pairs lie closer to the classification line than the star-forming field galaxies.
\label{class_diag}}
\end{figure}

In Figure~\ref{class_diag}, we compare the position of the galaxy pair members with the field galaxies on the standard optical diagnostic diagrams.  These diagrams are based on strong optical line ratios, and are commonly used to separate AGN from star-forming galaxies \citep{Baldwin81,Veilleux87,Kewley01b}.  Galaxies dominated by star-formation lie below the theoretical classification line derived by \citet{Kewley01a}.   
The position of the star forming galaxies on these diagrams is determined primarily by (1) metallicity, (2) the ionization state of the gas, and (3) star-formation history \citep{Kewley01a}.  
Figure~\ref{class_diag} shows that the star-forming galaxy pairs lie closer to the classification line than the field galaxies.  
 \citet{Donzelli00} found a similar result when they compared merging with interacting galaxies on the diagnostic diagrams.  They found that star-forming mergers occupy a higher locus on the diagnostic  diagrams than well-separated, interacting galaxies.  Donzelli et al. interpret this result as a consequence of a greater level of excitation in the nuclear spectra of galaxy mergers and they suggest that the greater level of excitation could result from a lower gas metallicity in the mergers.    High excitation is correlated with low gas metallicity;  the fewer metals in a galaxy, the fewer coolants are available and the gas maintains a high temperature.  The high temperature results in reduced recombination rates, and therefore higher excitation lines like \OIII\ are produced.  
However,  \citet{Marquez02}  find that the nuclear \NIIHa\ ratios for 13 non-disrupted galaxy pairs are similar to those of isolated galaxies.   We emphasize that the position on the diagnostic diagrams is not a simple function of metallicity, and that the pairs locus may result from one or a combination of a lower metallicity, higher ionization state, or a long (continuous) star formation history lasting $\geq 6$~Myr \citep{Kewley01a}.    In Sections~\ref{ionization} and~\ref{metallicity} we directly compare the ionization, metallicity and star formation properties of the galaxy pairs with the field galaxy sample.

\epsscale{1.2}

\section{Ionization Parameter}\label{ionization}

The ionization state of the gas is commonly described in terms of the local ionization parameter ($q$), defined as the number of hydrogen ionizing photons passing through a unit area  per unit density of hydrogen:

\begin{equation}
q=\frac{S_{{\rm H}^{0}}}{n}  \label{1}
\end{equation}
where $S_{{\rm H}^{0}}$ is the ionizing photon flux through a unit 
area, and $n$
is the local number density of hydrogen atoms. The ionization parameter $q$ has units of 
velocity (cm/s) and  can be physically
interpreted as the maximum velocity of an ionization front that can be
driven by the local radiation field.  Dividing by the speed of light gives the dimensionless local 
ionization parameter ${\cal U}\equiv q/c.$   

\begin{figure}
\plotone{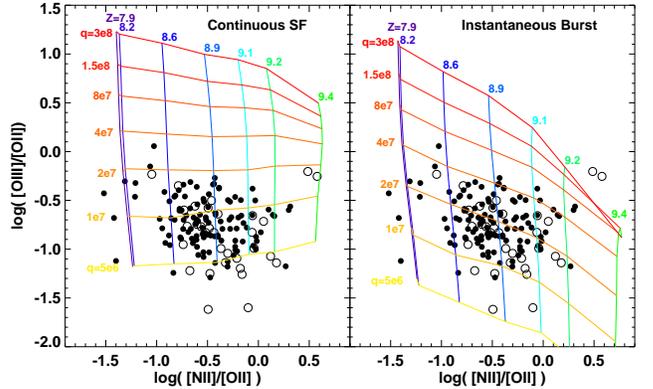}
\caption{The \OIII~$\lambda 5007$/\OII~$\lambda 3727$ ratio versus the \NII~$\lambda 6584$/\OII~$\lambda 3727$ ratio diagnostic diagram.  The galaxy pairs sample and the NFGS sample (with $cz \geq 2300$~km/s) are plotted as filled and unfilled circles respectively.  The grids show the locus of 
the theoretical star-forming models of \citet{Kewley01a} for two extreme burst scenarios: (left) continuous star formation, and (right) instantaneous zero-age burst.    The 
metallicity $Z$ is defined in terms of \OH\ where the currently accepted value of solar is 8.69+/-0.05 dex 
\citep{Allende01}.  The ionization parameter $q$ is in units of $cm/s$.
The galaxy pairs sample contains more galaxies at relatively low metallicities than the NFGS.  The galaxy pairs have higher \OIIIOII\ ratios on average than the field galaxies, indicating either a higher mean ionization parameter, a longer episode of star formation, or a combination of the two.
\label{OIIIOII_ratio}}
\end{figure}

Many of the common optical metallicity diagnostics are sensitive to
the ionization parameter.  So long as the EUV spectrum
of the exciting source is reasonably well constrained, the ionization
parameter can be determined using the ratios of emission-lines of
different ionization stages of the same element.  The most commonly used
ionization parameter diagnostic is based on the ratio of the \OIII~$\lambda 5007$ and
\OII~$\lambda 3727$ emission line fluxes \citep[e.g.,][]{Kewley02a}.   Figure~\ref{OIIIOII_ratio} 
shows the \OIIIOII\ vs \NIIOII\ line diagnostic diagram and the model grids of \citet{Kewley01a}.   The 
metallicity $Z$ is defined in terms of \OH\ where the currently accepted value of solar is $8.69 \pm 0.05$  \citep{Allende01}.
Kewley et al. calculated these grids using a combination of the P\'{e}gase stellar population synthesis  code \citep{Fioc97} and the Mappings photoionization code \citep{Sutherland93,Groves04b}.  
The models show that the position of galaxies on this diagram is determined by their ionization parameter, metallicity and star formation history.  We examine two star formation history models:
(a) a continuous star formation model, and (b) an instantaneous zero age burst model.   
The galaxy pairs are shifted towards a greater \OIIIOII\ ratio in Figure~\ref{OIIIOII_ratio} than the field galaxies.  Their position on these diagrams may result from either a greater
ionization parameter, a continuous star formation history, or a combination of the two.  A greater
ionization parameter could result from a greater number of ionizing photons per unit area or a lower
hydrogen density.  More ionizing photons may result from a larger fraction of young, 
hot stars within the nuclear regions in the galaxy pairs compared with the field galaxies.   This picture 
is consistent with numerous studies indicating that nuclear star formation increases in galaxy 
interactions and mergers \citep[e.g.,][and references therein]{Bushouse88,Donzelli97,Barton03,Nikolic04}.  We note that shock excitation cannot explain 
the line ratios observed in Figures~\ref{OIIIOII_ratio} and~\ref{class_diag}.  Shock excitation raises the \OIIIOII\ ratio, but it also raises the \SIIHa\ ratio by a similar amount \citep[e.g.,][]{Sutherland93,Kewley01b}; we do not observe greater \SIIHa\ ratios for the galaxy pairs.

\begin{figure}
\plotone{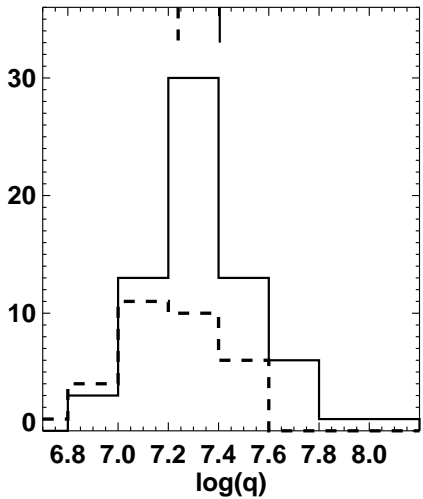}
\caption{The ionization parameter distribution for the galaxy pairs (solid line) and the NFGS (dashed line), assuming an instantaneous burst model.  The logarithm of the mean ionization parameter for each distribution is shown at the top.   The logarithm of the mean ionization parameter is $\sim 7.2$ for the NFGS and $~\sim 7.4$ for the pairs sample, corresponding to a difference of $\sim 8\times 10^{6}$~cm/s.
\label{q_pairs_NFGS}}
\end{figure}

We calculate the ionization parameter using the \OIIIOII\ diagnostic in \citet{Kobulnicky04} using the metallicity estimates described in Section~\ref{metallicity}.  The Kobulnicky \& Kewley  \OIIIOII\ diagnostic utilizes the instantaneous burst models shown in Figure~\ref{OIIIOII_ratio}.   Figure~\ref{q_pairs_NFGS} shows the ionization parameter distribution for each sample assuming the instantaneous burst model.  The logarithm of the mean ionization parameter is $\sim 7.2$ for the NFGS and $\sim 7.4$ for our
pairs sample, corresponding to a difference of $\sim 8\times 10^{6}$~cm/s.  Relative errors between 
ionization parameters estimated using the same calibration are typically $<< 0.1$~dex.
If continuous burst models 
are used, our ionization parameter estimates for the NFGS and the pairs are lowered by 
$\sim 0.3$~dex, but the difference between the two samples remains $\sim 8\times 10^{6}$~cm/s.
We emphasize that it is not possible to determine whether differences in ionization parameter, burst
scenario, or a combination of the two is responsible for the \OIIIOII\ difference between the pairs and the field galaxies in Figure~\ref{OIIIOII_ratio}. 

The recent star formation history of the galaxy pairs sample was investigated by \citet[][; hereafter BGK03]{Barton03}.
They showed that the equivalent width (EW) of \Ha\ and other emission-lines correlate strongly with 
the inverse of the pair spatial separation ($\Delta D$) or velocity separation.  BGK03 used EW(\Ha) with stellar population synthesis models to estimate the time since the most recent burst began, or the "burst timescale".  By matching the burst timescale with the dynamical timescale, BGK03 obtain a merger
scenario in which the first close pass initiates a starburst, increasing EW(\Ha).   BGK03 note that this  scenario is supported by the pairs data as long as the triggered bursts last longer than $\sim 10^8$~years.  This timescale for star formation is consistent with our continuous star formation model in Figure~\ref{OIIIOII_ratio}, suggesting that the star formation history may contribute to the increased 
\OIIIOII\ ratio seen in the galaxy pairs.

\epsscale{1.0}


Figure~\ref{OIIIOII_ratio} shows that the metallicity predicted by the theoretical models is independent 
of burst scenario.  The metallicity range spanned by the galaxy pairs is larger 
than the field galaxies; the pair sample contains more galaxies with relatively low metallicities
(\OH$<8.8$).   
Note that the \NIIOII\ ratio becomes insensitive to metallicity at low \NIIOII\ ratios.  In the following section we derive the metallicities for each sample and we investigate the cause of the metallicity distribution in Figure~\ref{OIIIOII_ratio}.

\section{Metallicity and Star Formation in Pairs}\label{metallicity}

We calculate metallicities for the pairs and the field galaxies using the strong optical emission-line ratios.   
There is a well-known discrepancy between metallicities calculated using 
theoretical strong-line diagnostics derived from photoionization models 
\citep[e.g.,][and many others]{Mcgaugh91,Zaritsky94,Kewley02b}, 
diagnostics based on electron-temperature measurements 
\citep[e.g.,][]{Kewley05b,Garnett04,Garnett04b,Bresolin04}, and diagnostics based on 
metallicities derived from a combination of the two methods \citep[e.g.][]{Pettini04,Denicolo02}.
This discrepancy manifests itself as a systematic offset with a maximum difference 
of $\sim 0.4$~dex in \OH\ for star forming galaxies.   
The cause of this offset is unknown \citep[see][for a discussion]{Garnett04,Garnett04b,Stasinska05}.  Until this discrepancy is resolved, 
absolute metallicities derived using any method should be treated with caution.  
Fortunately, the difference between the various diagnostics is systematic.  Therefore
the error in {\it relative} metallicities derived using {\it the same} method, 
regardless of which method is used, is $<< 0.1$~dex \citep{Kewley05b}.  
We therefore use metallicities derived using the same method 
{\it only for relative comparisons} between samples or among galaxies within a 
sample.  

We use a combination of the \citet{Kewley02b} ``recommended'' method (hereafter KD02) 
and the \citet{Kobulnicky04} recalibration of the 
\citet{Kewley02b} \NIIHa-metallicity and \R23-metallicity relations.  These methods are based on a self-consistent 
combination of detailed current stellar population synthesis and photoionization models.   
These models successfully reproduce the \HII\ region abundance sequence 
\citep{Dopita00}, the abundance sequence of star-forming galaxies \citep{Kewley01a}, 
and individual star clusters within galaxies \citep{Calzetti04}.
Our final metallicities are determined preferentially according to the following criteria:

\begin{enumerate}
\item If \NII~$\lambda 6584$, \OII~$\lambda3727$, \Ha, and \Hb\ are available, and 
log(\NIIOII)$\gtrsim -1.0$ we  use the Balmer decrement to correct the \NIIOII\ ratio for 
reddening and we apply the KD02 \NIIOII\ diagnostic to determine metallicities.   The
\NIIOII\ ratio is extremely sensitive to metallicity but is insensitive to ionization parameter
variations for log(\NIIOII)$\gtrsim -1.0$.  

\item If  \OII~$\lambda3727$, \OIII~$\lambda5007$, \Ha, and \Hb\ are available, and 
 log(\NIIOII)$\lesssim -1.0$, we use the \citet{Kobulnicky04} recalibration of the 
\citet{Kewley02b} \R23 -metallicity relation.

\item If \NII~$\lambda 6584$, and \Ha\ are available but \OIII~$\lambda5007$ 
and/or \Hb\ are not available we use the average of the \citet{Kobulnicky04} recalibration of the 
\citet{Kewley02b} \NIIHa -metallicity relation and the \citet{Charlot01} Case A \NIISII\ calibration.  
For high metallicities (\NIIHa$> 0.4$) where \NIIHa\ becomes insensitive to metallicity  we use 
the \NIISII\ calibration alone.  We assume that the ionization parameter 
for these galaxies is the same as the sample mean ionization parameter.  The mean 
ionization parameter is determined iteratively using the reddening-corrected \OIIIOII\ 
ratio as described in KD02.  Deviation from the mean ionization parameter value 
is the dominant error source for metallicities calculated using \NIIHa\ and \NIISII .

\end{enumerate}

A total of 101/105 \HII\ region-like galaxies in the NFGS and 153/153 pair members 
satisfy these criteria.  We investigate the effect of residual AGN contribution on the derived 
metallicities in the Appendix.

\subsection{Luminosity-Metallicity Relation for Galaxy Pairs }\label{LZ}

\begin{figure}
\plotone{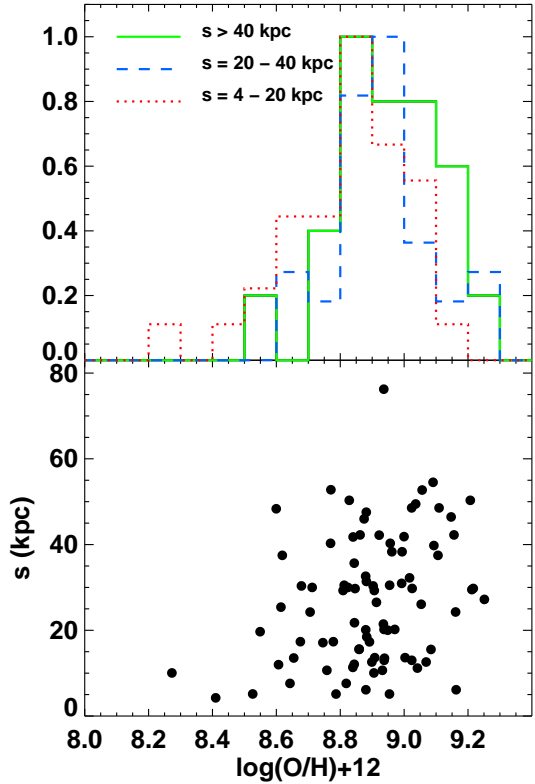}
\caption{(a) The normalized metallicity distribution for the pair members with 
projected separations $4 < s \leq 20$~ kpc/h (red dotted line), $20 < s \leq 40$~ kpc/h (green solid line),
$40 < s$~kpc/h (blue dashed line), (b) the projected separation versus the metallicity for the pair members.  The lowest metallicities occur predominantly in the close pairs.
\label{Abund_sep}}
\end{figure}

In Figure~\ref{Abund_sep}a we show the metallicity distribution for the pair members with 
projected separations $4 < s \leq 20$~ kpc/h (red dotted line), $20 < s \leq 40$~ kpc/h (green solid line),
$40 < s$~kpc/h (blue dashed line).    Figure~\ref{Abund_sep}b plots the projected separation versus the 
metallicity.   Galaxies with projected separations $<20$~kpc/h (close pairs) have a smaller mean metallicity ($8.85 \pm 0.02$\footnote[1]{Error is the standard error of the mean}) than galaxies with larger projected separations ($8.97\pm 0.02$).

\epsscale{1.2}
\begin{figure}
\plotone{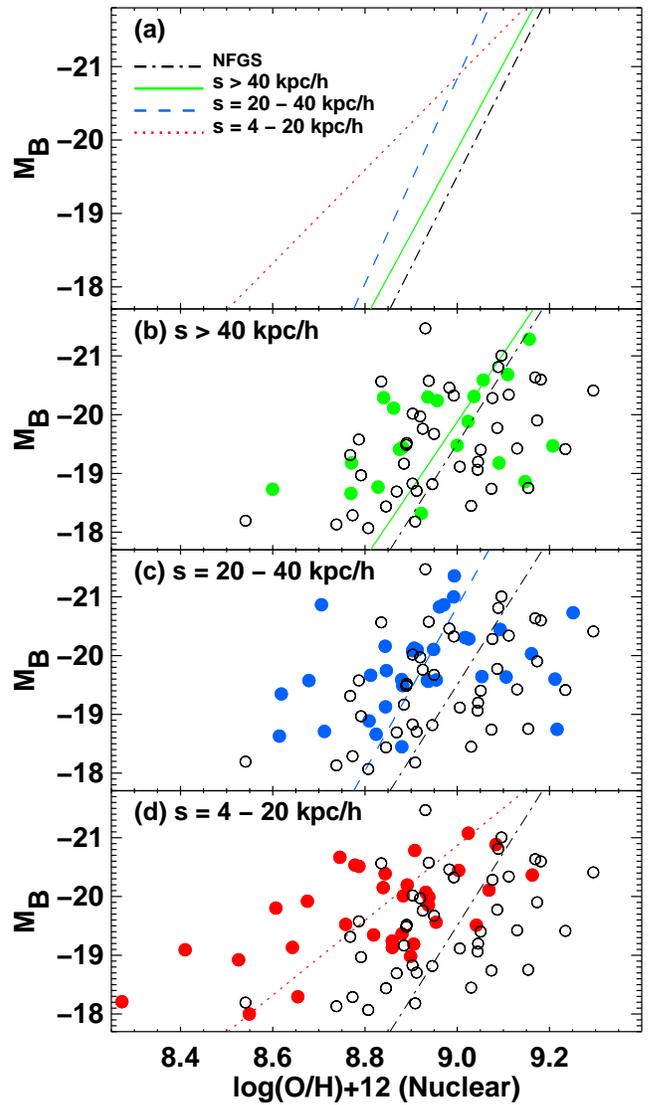}
\caption{(a) The least squares best-fit to the B-band luminosity metallicity relations for the field galaxies (dot-dashed line), and the pair members with 
projected separations $4 < s \leq 20$~ kpc/h (red dotted line), $20 < s \leq 40$~ kpc/h (green solid line),
$40 < s$~kpc/h (blue dashed line), (b-c) comparison between the luminosity-metallicity relation of 
the field galaxies (unfilled circles) and the galaxy pairs (filled circles) for the three groups of 
projected separation in (a).  The close pairs have a luminosity-metallicity relation shifted 
towards lower metallicities than the field galaxies or than the more widely separated pairs.
\label{LZplot}}
\end{figure}

\epsscale{1.0}
\begin{figure}
\plotone{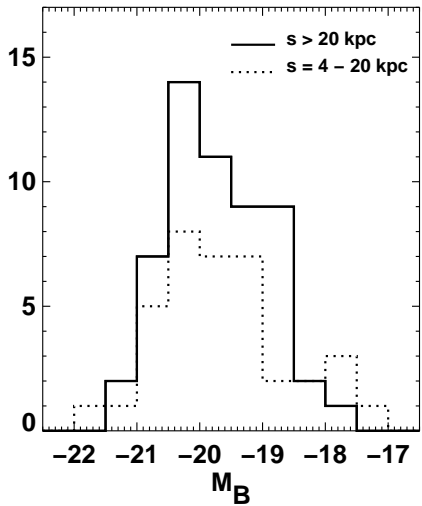}
\caption{The luminosity distribution for the galaxy pair members with 
projected separations $4 < s \leq 20$~ kpc/h (dotted line), and $s > 20$~ kpc/h (solid line).   The 
Kolmogorov-Smirnov two-sample test indicates that the two samples are consistent with a single distribution at the 72\% confidence level.  
\label{sep_lum_hist}}
\end{figure}

The metallicity distribution in Figure~\ref{Abund_sep}a for the close pairs is asymmetric and skewed towards lower metallicities compared to the more widely separated pairs.   This difference is reflected 
in the luminosity-metallicity relation.  In Figure~\ref{LZplot}  we 
plot the luminosity-metallicity relation for the field galaxies (unfilled circles) and the pair members (filled 
circles) for the three ranges of projected separation.  Panel (a) shows the least-squares best-fit to 
the L-Z relation for the field galaxies (dot-dashed line), and to the pairs in three groups of projected separation: $4 < s \leq 20$~ kpc/h (red dotted line), $20 < s \leq 40$~ kpc/h (green solid line), $40 < s$~kpc/h (blue dashed line).   Panels (b-d) compare the L-Z relation for the pairs for each of the three pair groups with the L-Z relation for the field galaxy sample.  Excluding the one outlier, the pairs with $s>40$~kpc/h and with 
$20 < s \leq 40$~ kpc/h have similar L-Z relations to the field galaxies.   This result is consistent with \citet{Marquez02} who find that the nuclear \NIIHa\ ratios for 13 non-disrupted galaxy pairs are similar to those of isolated galaxies; many of the Marquez et al. non-disrupted pairs are widely separated.

Figure~\ref{LZplot}d shows a striking difference; the close pairs have an L-Z relation shifted toward lower metallicities.    For galaxies at the same luminosity, close pairs have lower metallicities than the field galaxies.  Even more striking, there are 10 close pairs with metallicities lower than any field galaxy at the same luminosity.     R-band luminosities produce identical results.

\begin{figure}
\plotone{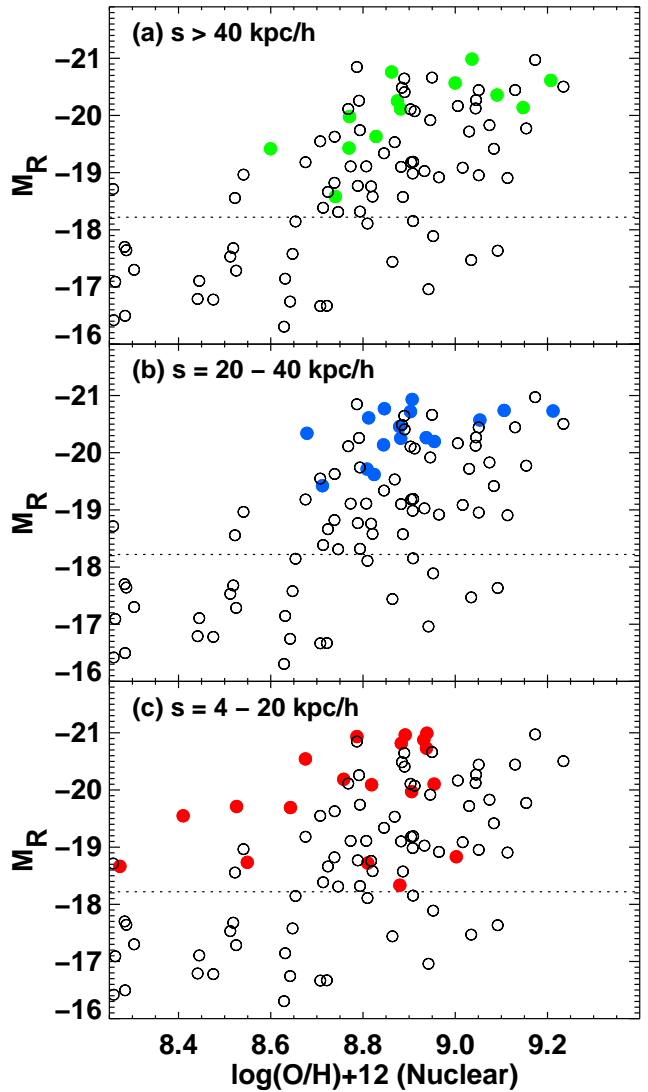}
\caption{(a) The R-band luminosity-metallicity relation of 
the field galaxies (unfilled circles) and the galaxy pairs (filled circles) for the three groups of 
projected separation as shown.  The R-band luminosity for the pairs has been corrected for the 
strength of the recent central burst of star formation \citep{Barton03}.  The dotted line shows the detection limit of the pairs sample imposed by the recessional velocity selection criterion ($cz>2300$~km/s).   While we must compare metallicities of galaxies at the same luminosity, we show the position of the field galaxies with $cz\leq2300$~km/s  for comparison.  The close pairs have a luminosity-metallicity relation that is shifted towards lower metallicities than the field galaxies or than the more widely separated pairs, even after correction for the recent bursts of central star formation.
\label{LZ_Rmag}}
\end{figure}

This L-Z shift does not appear to 
result from a difference in luminosity distribution between the three projected separation groups.  We compare the luminosity distribution for the whole star-forming pairs sample (i.e. without luminosity cuts) in Figure~\ref{sep_lum_hist} .  The close pairs (dotted line) do not lie at significantly higher luminosities than the more widely separated pairs (solid line).     The Kolmogorov-Smirnov two-sample test indicates that the two samples are consistent with a single distribution function at the 72\% confidence level. BGK00 calculated the central burst strength $s_{R}(t)$ of the galaxy pairs by fitting the observed B-R and \Ha\ colors with stellar population synthesis models.   Burst strength provides a measure of the current fraction of the central R-band flux 
originating from a new burst of star formation; pair members with recent bursts will have large central burst strengths.   A total of 59 galaxies in our pairs sample have measured central burst strengths \citep[][hereafter BGK03]{Barton03}.    The recent central bursts contribute $\sim 0.1$~dex in R-band on average, with a range of $0.004 -  0.3$~dex.   Because the central burst strength is a measure of the central R-band flux originating from recent star formation, we can subtract the recent central star formation from the total R-band luminosity with knowledge of the slit-to-total R-band flux ratio and ask whether we can recover a "normal" L-Z relation characteristic of the underlying galaxy.
Figure~\ref{LZ_Rmag} shows that the corrected total R-band L-Z relation has the same L-Z shift as in Figure~\ref{LZplot}.

\epsscale{1.2}

Figure~\ref{LZ_Rmag} is based on the assumption that the recently triggered star formation is located 
within the nucleus.  However, even if we were to assume that the central burst strength is equal to the recent burst strength across the entire galaxy measured by the global R-band light, the R-band L-Z relations for the pairs and the field galaxies still do not agree.   This exercise shows that the close pairs L-Z relation is not a result of a recent burst superimposed on a low luminosity, low metallicity galaxy.

Figures~\ref{LZplot} and~\ref{LZ_Rmag} show that the upper metallicity bound for the close pairs is lower than for the more widely separated pairs or the field galaxies.  The wide pairs subsample ($40 < s$~kpc/h) has 5/34 (15\%) galaxies with metallicities \OH$>9.1$.  This fraction is comparable to the fraction (4/21; 19\%) of intermediate pairs ($20 < s \leq 40$~ kpc/h)  and field galaxies (8/43; 19\%) with metallicities \OH$>9.1$. 
In contrast, the close pairs sample has only 1/37 (3\%) galaxies with  metallicities \OH$>9.1$.    Statistically, we would expect 6-7 galaxies in the close pairs sample to have \OH$>9.1$ if there is no shift in the upper metallicity bound with smaller projected separations.  The deficit of close pairs with  \OH$>9.1$ as well as the shift in luminosity-metallicity relations for the close pairs leads us to conclude that the lower metallicities observed in the galaxy pairs probably result from a change in the gas-phase metal fraction in the nuclei of at least some of the close pair members.    

As a final check, we investigate the possibility of extended triggered star formation in the galaxy pairs.
Extended star formation occurs in both starburst galaxies (\citet{Rieke85,Schweizer86}; see \citet{Schweizer05} for a review) and post-starburst systems such as E+A galaxies 
\citep{Franx93,Caldwell96,Norton01,Pracy05}.   Our spectra represent a luminosity-weighted mean of 
the individual star forming regions observed within the spectral aperture.  If star formation is systematically triggered at increasingly large radii as the interaction progresses, the extended regions may dominate the light in the spectroscopic aperture at the late merger stages.  In this case, our measured metallicities for the close pairs could represent the metallicity in the extended regions rather than the nuclear star forming regions.  If this process is important, we would expect to see an anti- correlation between aperture covering area and metallicity, particularly for the close pairs, in the sense that the lowest metallicities would be measured through the largest apertures.  Our pairs spectra were obtained through a fixed-width aperture of 3 arcseconds.   At the redshifts of our pairs sample, this aperture corresponds to a physical width of 0.01 - 2 kpc.   The length of the slit varies from galaxy to galaxy, and at the distance of our sample, corresponds to a length of 0.01 - 9 kpc, or 0.02 - 6.5 disk scale lengths.

\begin{figure}
\plotone{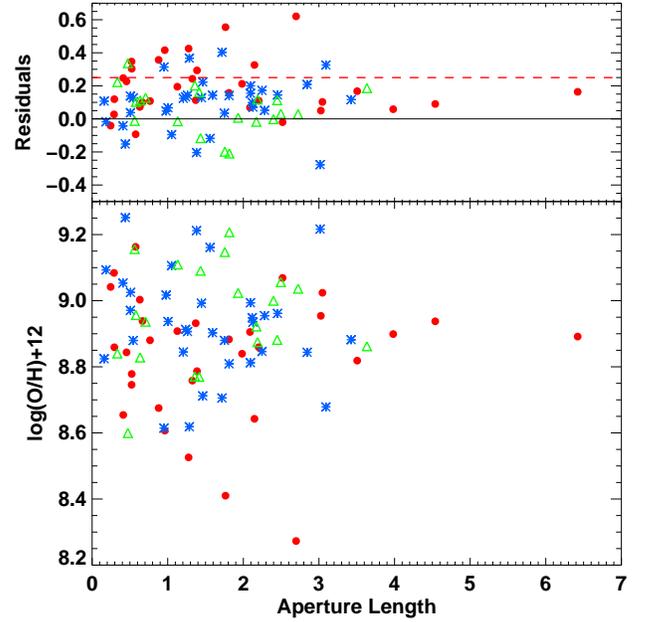}
\caption{Nuclear aperture length (units of disk scale length) versus the residuals from the field 
galaxy LZ-relation (dot-dashed line in Figure~\ref{LZplot}) for the galaxy pairs sample.  The low-metallicity residuals lie to the right of the red-dashed line.  The black line shows the position of the 
field galaxy LZ relation.  The galaxies with large residuals do not have the largest aperture lengths.  This result indicates that systematic triggering of starbursts at larger radii does not play a major role in the 
shift in LZ relation for close pairs. 
\label{Ap_Z}}
\end{figure}

Figure~\ref{Ap_Z} shows the relationship between aperture length in units of disk scale length at the distance of each galaxy as a function of their residual metallicities from the field galaxy LZ relation 
(dashed line in Figure~\ref{LZplot}).  We do not see an anti-correlation between metallicity and aperture length.  On the contrary, the lowest metallicities are measured through small-to-average sized aperture lengths (0.4 - 2.7 disk scale lengths, or 1-4~kpc).  Similar results are found for aperture width and total aperture size.  We conclude that the low metallicities in our close pairs probably do not result from star formation being systematically triggered at increasingly large radii as the interaction progresses.  

\epsscale{1.2}

 We investigate the L-Z shift in terms of morphological and spectroscopic properties in Appendix~\ref{Other_effect}.  None of these properties can explain the L-Z shift towards lower metallicities.  We tentatively conclude that the lower nuclear metallicities observed in the close galaxy pairs result from a change in the gas-phase metal fraction.   Because the lower-than-field metallicities occur predominantly in the close pairs, these lower metallicites may result from an interaction-induced process.   Interaction-induced gas infall from the outer, less enriched regions of close pairs initiated by the first close pass is one possibility.  Merger models predict that substantial gas inflows occur towards the nucleus of disk galaxies undergoing interactions \citep{Barnes96,Mihos96,Bekki01,Iono04,Springel05}.   
In Sections~\ref{bursts} and \ref{bulges} we investigate potential indicators of gas 
flows in our pairs sample.

\subsection{Tidally Triggered Central Bursts} \label{bursts}

Theory predicts that gas flows in galaxy interactions trigger central bursts of star formation 
\citep{Barnes96}.  This infall is predicted to begin during the first close pass.   As the galaxies move apart,  EW(\Ha) decreases as 
the new population raises the continuum around \Ha.   Our sample may therefore select against galaxies that are enriched from the recent burst of star formation at large projected separations.  

\begin{figure}
\plotone{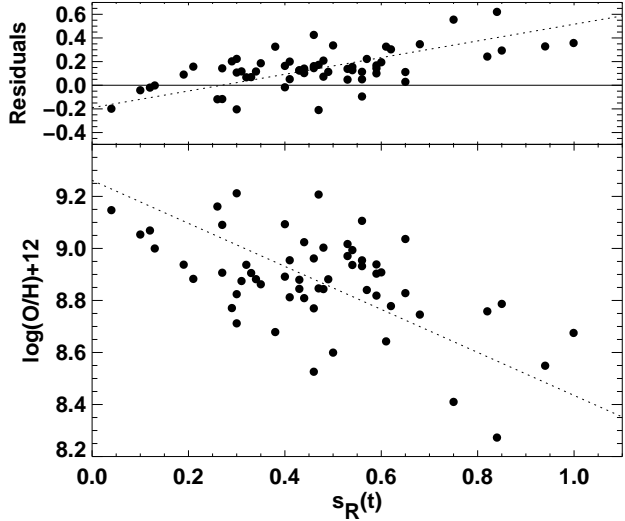}
\caption{Bottom Panel: Nuclear Metallicity versus central burst strength ($s_{R}(t)$) for the galaxy pairs.  The dotted line shows the least squares line of best-fit to the data.  The metallicity and central burst strength are anti-correlated.  Top Panel: Residuals from the field galaxy LZ relation versus central burst strength.  The strong correlation between LZ residuals and central burst strength indicates that the anti-correlation between metallicity and central burst strength is stronger than the strength that would result from a simple luminosity-star-formation and luminosity-metallicity scaling relationship.
\label{Abund_burst_str}}
\end{figure}

We plot the nuclear metallicity versus the central burst strength in the lower panel of Figure~\ref{Abund_burst_str}.  The 
dotted line shows the least squares line of best-fit to the data.  The metallicity and burst strength are
anti-correlated.  The Spearman Rank correlation test gives a correlation coefficient of -0.43. The two-sided probability of finding a value of -0.43 by chance is 0.1\%.    
If gas inflows have triggered the nuclear bursts, then those gas inflows may have carried more pristine 
gas from the outskirts of the galaxy into the nuclear regions, diluting the central metallicity.  
BGK03 showed that strong central bursts tend to occur at close projected separations.
Gas infall is expected to occur after two galaxies undergo their the first close encounter \citep{Barnes02,Barnes96}.  
 
 The top panel of Figure~\ref{Abund_burst_str} shows that the pair member residuals from the mean field galaxy LZ relation (dot-dashed line in Figure~\ref{LZplot}) are correlated with the central burst strength.  
Residuals are calculated as the shortest distance to the mean field galaxy LZ relation; positive residuals correspond to negative metallicities. 
 The Spearman Rank correlation test gives a correlation coefficient of 0.52. The two-sided probability of finding a value of 0.52 by chance is 0.007\%.   This result indicates that the anti-correlation between 
burst strength and metallicity is over and above what one might expect from a normal luminosity-metallicity scaling relationship.

\begin{figure}
\plotone{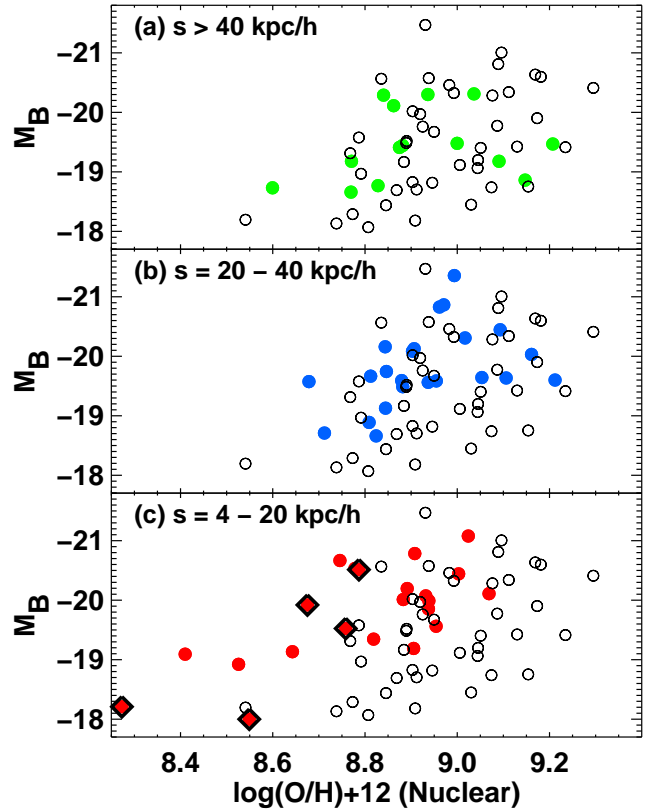}
\caption{(a) The luminosity-metallicity relation for pair members with measured central burst 
strengths and projected separations between (a) $4 < s \leq 20$~ kpc/h, (b) $20 < s \leq 40$~ kpc/h, and (c) $40 < s$~kpc/h.   The five galaxies with strong central bursts (diamonds; $s_{R}(t)>0.8$) are all in close pairs with metallicities lower or at the lower bound of the field galaxies.  This result provides circumstantial evidence that gas flows have been important in the low metallicity close pairs.
\label{LZ_burst_str}}
\end{figure}

In Figure~\ref{LZ_burst_str}, we show the luminosity-metallicity relation for the pairs with central burst strengths measured by BGK00 (solid symbols).  We show the field galaxies
for comparison (unfilled circles).
BGK00 identified six galaxies in the pairs sample that have high central burst strengths indicative of 
strong tidally triggered bursts ($s_{R} (t) > 0.8$).  Of these six galaxies, five have luminosities between $-21.5 \leq {\rm M_{B}} \leq -18$ and are dominated by star formation (diamonds in  Figure~\ref{LZ_burst_str}).   All five galaxies have a companion at close projected separation $s<20$~kpc/h.  Their metallicities are all less than or equal to the lowest metallicity field galaxies at the same luminosity.  The one remaining galaxy with a high central burst strength that does not satisfy our selection criteria is classified as 'Ambiguous'.  This galaxy may be a composite so we have excluded it from our study.  We note that 
this object is also in a close pair with $s<20$~kpc/h, has a metallicity  of \OH$=8.63$ and luminosity ${\rm M_{B}}=-19.40$, placing it firmly within the low metallicity region occupied by the five galaxies shown in Figure~\ref{LZ}.    The fact that all of the high central burst strength galaxies have both close 
companions and predominantly lower metallicities than the field galaxies provides strong circumstantial evidence that gas infall has played an important role in these low central metallicity galaxies.

\subsection{Blue Bulges}\label{bulges}

Recent observations suggest that some bulges form within pre-existing disk galaxies \citep{Kannappan04}.   These bulges within disks may be linked to disk 
gas inflow and central star formation triggered by either internal processes or galaxy mergers and 
interactions \citep{Tissera02}.   Active growth of bulges within disks can be identified using $B-R$ color profiles; galaxies which are actively growing bulges are likely to have bluer colors within their half-light radii compared to the colors of their outer disks \citep[][; BGK03]{Kannappan04}.   We refer to bulges selected using 
$B-R$ colors as "blue bulges".

BKG03 showed that the $B-R$ color profiles of the six galaxies with strong central bursts discussed in Section~\ref{bursts}  also have blue central dips.   Here we investigate whether the presence of blue bulges is correlated with metallicity.    The $B-R$ colors for the inner and outer disks are available for 87 
of our galaxy pair members.

\begin{figure}
\plotone{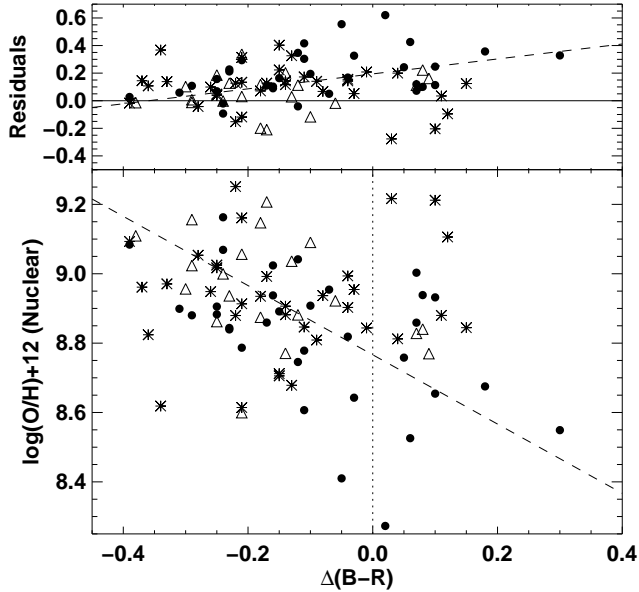}
\caption{Lower Panel: Nuclear Metallicity versus $\Delta(B-R)$ for the galaxy pairs where $\Delta(B-R)$ equals 
the outer disk (75\%-light radius) $(B-R)$ color minus the $(B-R$ color in the central half-light radius.
Galaxies lying above the dotted line are likely to have blue bulges \citep{Kannappan04}.  The dashed line is the least-squares line of best fit to the sample.  Upper panel:  Residuals from the field galaxy LZ relation versus  $\Delta(B-R)$ for the galaxy pairs.  The LZ residuals are weakly correlated with $\Delta(B-R)$.
\label{dB-R}}
\end{figure}

In Figure~\ref{dB-R}, lower panel we plot metallicity versus $\Delta(B-R)$ where $\Delta(B-R)$ is the 
difference between the outer disk (75\%-light radius) color and the color inside the half-light radius 
($r_e$).  The different symbols correspond to the three pairs groups with projected separation 
$4 < s \leq 20$~ kpc/h (circles), $20 < s \leq 40$~ kpc/h (asterixes), $40 < s$~kpc/h (triangles).
Metallicity and $\Delta(B-R)$ are weakly anti-correlated; the Spearman Rank correlation 
coefficient is -0.29 and the probability of obtaining this value by chance is 0.7\%.   This correlation is 
dominated by the close pairs with low metallicities.

Kannappan et al. classify galaxies with blue bulges as those galaxies with $\Delta(B-R)>0$ (dotted line in Figure~\ref{dB-R}).  With this definition we see that galaxies with blue bulges cover the full range of metallicities in the pairs sample.  However, galaxies with low metallicities are a more likely to have blue bulges compared to higher metallicity galaxies; for metallicities below \OH$<8.8$, 33\% of galaxies have blue bulges; for metallicities \OH$>8.8$, 19\% have blue bulges.    In the blue bulge sample ($\Delta(B-R)>0$), a larger fraction of galaxies are in close pairs compared to the 
remainder of the sample (53\% c.f. 34\%).

The top panel of  Figure~\ref{dB-R}, shows that there is a weak correlation between the field galaxy LZ residuals and $\Delta(B-R)$.  The Spearman Rank coefficient is 0.22 with a probability of obtaining this value by chance of 4\%.

\begin{figure}
\plotone{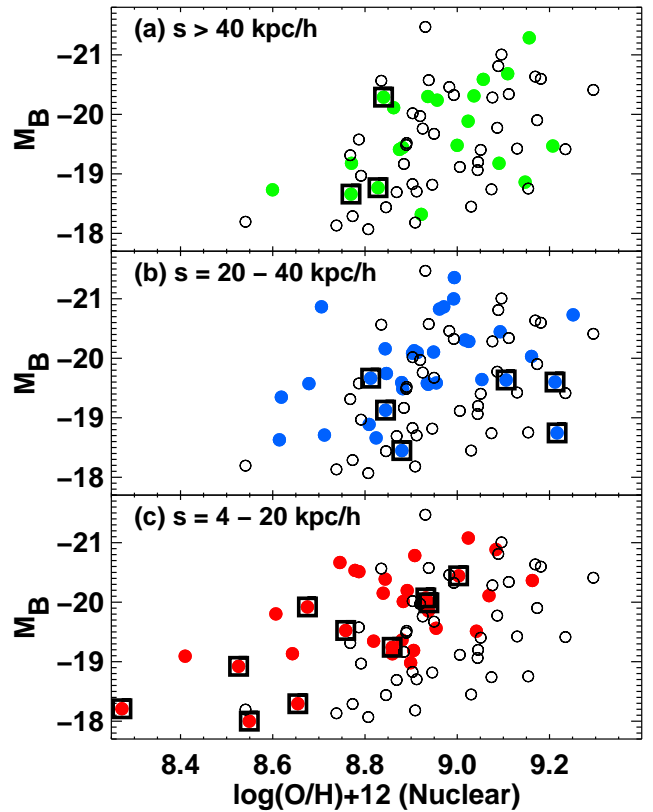}
\caption{(a) The luminosity-metallicity relation for pair members with measured  $\Delta(B-R)$  and projected separations between (a) $4 < s \leq 20$~ kpc/h, (b) $20 < s \leq 40$~ kpc/h, and (c) $40 < s$~kpc/h.  Pair members with blue bulges ($\Delta(B-R)>0$) are shown as squares.  Although most of 
the close pairs with lower metallicities than the field galaxies have blue bulges, galaxies with blue bulges occupy the full range of projected separation and metallicity in the pairs sample.
\label{LZ_BB}}
\end{figure}

In Figure~\ref{LZ_BB} we show the luminosity-metallicity relation for the three ranges in projected 
separation.  Galaxies with blue bulges  are shown with a large square outline.   Although most of 
the close pairs with lower metallicities than the field galaxies have blue bulges, galaxies with blue bulges occupy the full range of luminosity, projected separation, and metallicity in the pairs sample.

\section{Discussion}  \label{discussion}

Our close pairs have lower metallicities, higher central bursts strengths and are more likely to be have blue bulges than galaxy pairs at wider separations.  These low metallicity pairs are probably at a later stage in the merger process than pairs with higher metallicities.

Smoothed particle hydrodynamic merger simulations of \citet{Barnes96} and \citet{Mihos96} predict that a galaxy interaction influences the gas flow within a galaxy in two ways: (1) the tidal forces directly affect the gas by inducing gravitational torques which remove angular momentum from the gas, and (2) the gravitational influence of the companion deforms the distribution of the stars and gas within the galaxy.  The distorted stellar distribution changes the gravitational potential of the galaxy, which then affects the motion of the gas.   The gas rapidly accumulates into the 
stellar arms that subsequently form shocked, dense filaments.  In a dissipative medium, these filaments allow substantial streaming of gas towards the nucleus.  The detailed theoretical response of the gas in merging galaxies was recently investigated by \citet{Iono04}.  Iono et al. predict that the average gas inflow rate to the central $1 - 2 $~kpc is $\sim 7$~\Msun/yr with a peak infall rate of $\sim 17$~\Msun/yr.  These simulations predict that the gas flows towards the central regions within 100 Myr after the initial collision, but before the two disks merge.  Evidence for ionized gas flows in interacting galaxies has recently been found in \Ha\ velocity maps of two galaxy pairs \citep{Rampazzo05}, and in radial velocity curves in the interacting system Arp 194 by \citet{Marziani03}.    Cold gas flows have been observed in a number of merging galaxies, including a galaxy pair separated by 29~kpc \citep{Chen02}.

BGK03 showed that the equivalent width (EW) of \Ha\ and other emission-lines correlate strongly with the inverse of the pair spatial separation or velocity separation.  By matching the burst timescale with the dynamical timescale, BGK03 obtain an interaction scenario in which the first close pass drives gas towards the central regions.  This gas infall initiates a central starburst which increases EW(\Ha).     As the burst ages and the projected separation increases, EW(\Ha) decreases as the new population raises the continuum around \Ha\ and enriches the gas.   BGK03 note that this scenario is supported by the pairs data as long as the triggered bursts last longer than $\sim 10^8$~years.  

If a galaxy pair member is a late-type spiral with a strong metallicity gradient, then gas infall initiated by the first close pass could move less enriched gas from the outskirts of the galaxy into the central regions covered by our aperture.   Massive late-type spirals like M101 have large 
metallicity gradients; the \HII\ region metallicities decrease by an order 
of magnitude from the inner to the outer disk 
\citep[see e.g.,][for a review]{Shields90}.  Metallicity 
gradients may result from one or a combination of: (1) radial variation of stellar yields caused by IMF 
differences between the spiral arms and the interarm regions \citep[e.g.][]{Guesten82}; (2) a star formation rate dependence on radius \citep{Phillipps91}; (3) radial infall of primordial gas during disk formation \citep{Matteucci89,Pagel89,Edmunds95}.  Current chemical evolution models
include a combination of these three processes \citep[e.g.,][]{Churches01}.

Not all galaxies have strong metallicity gradients, and the lack of a gradient is often cited 
as evidence for radial gas flows \citep[see][for a review]{Henry99}.  
Many barred spiral galaxies have weaker metallicity gradients than 
spirals of similar type \citep[e.g.,][]{Pagel79,Roy97}, suggesting that 
radial gas flows suppress or change metallicity gradients 
\citep[e.g.,][]{Roberts79,Martin94,Roy97} \citep[but c.f.][]{Considere00}.  

We can compare the fraction of infalling gas required to produce a shift in our metallicities with those predicted by hydrodynamic merger models.
The nuclear metallicity of our close pairs is $\sim 0.2$~dex lower than the metallicity of pairs at larger separations or field galaxies at the same luminosity.   At the mean redshift of our sample, our nuclear spectral aperture corresponds to the central $\sim 1.5$ kpc.  Late-type spiral galaxies typically have metallicity gradients of $\sim 1$~dex from their nuclear to outer (8-9~kpc) \HII\ regions \citep[see][for a review]{Henry99}.  A $0.2$~dex reduction in the nuclear metallicity in our pairs sample requires the gas within the central 1.5~kpc to be diluted 
by 50-60\% with gas from the outer disk regions.   
Assuming a large initial central gas mass of $10^{9}$~\Msun \citep{Sakamoto99,Regan01}, and an infall rate of 7~\Msun/yr, to dilute the central gas mass fraction by 60\% requires gas to infall over a timescale of $\sim 9 \times 10^{7}$ years.  This infall timescale is consistent with the $1\times10^{8}$~Myr timescale for gas infall predicted by the merger models. 

With a lower initial central gas mass of $1\times 10^{8}$\Msun, enough gas  could accumulate from the outer regions to account for a 0.2~dex reduction in metallicity in only $9\times 10^{6}$ years.   The amount of dilution required is easily attainable in current merger simulations; merger models predict that $\sim 60$\% of the gas initially distributed throughout the progenitor disks may be driven into a small nuclear 'cloud' with dimensions of $\sim 0.1$~kpc during the merging process \citep{Barnes91,Barnes96,Barnes02}.  Evidence for such  gas redistribution has been found by \citet{Hibbard96}.  Hibbard et al. observed the distribution of cold and warm gas in galaxies along the Toomre sequence.  They conclude that \HI\ disks become more disrupted as the merger progresses, from 60\% of the \HI\ mass being found in the main disks and bulges in each galaxy in the early merger stages to nearly 0\% in the late merger stage.
 
If the lower metallicities in our close pairs are a consequence of gas infall, these galaxies must be at a late stage in the infall process but at an early stage for chemical enrichment by triggered star formation.  At $9.4 \times 10^{7}$ years, the merger simulations of \citet{Iono04} predict that the outer disk is no longer able to supply sufficient gas for infall, and the infall rate decreases. 
Observations of the metallicity gradients and the nuclear morphology in our close pairs could test whether the infall scenario described above is correct.

We note that modeling gas flows in galaxy interactions is subject to uncertainties in our understanding of the kinematic behaviour of the gas.  For example, a clumpy interstellar medium (ISM) may 
behave very differently in an interaction than the highly dissipative ISM that is traditionally assumed \citep{Charmandaris00}.

\section{Conclusions }\label{conclusions}

We compare the nuclear metallicity and ionization properties of a large objectively selected sample of galaxy pairs and N-tuples with those of the objectively selected Nearby Field Galaxy Survey.    We 
derive the first metallicity-luminosity relation for galaxy pairs and we compare this relation 
with the field galaxy luminosity-metallicy relation.  We find that:

\begin{enumerate}
\item  Galaxy pairs with close projected separations $s<20$~kpc/h have systematically lower 
metallicities ($\sim 0.2$~dex) than either the field galaxies or the widely separated pairs.   There are 10 close pairs with metallicities lower than any field galaxy at the same luminosity.    
\item We find a tight anti-correlation between nuclear metallicity and central burst strength in the galaxy 
pairs; lower metallicity pairs have greater central burst strengths.  This anti-correlation is stronger than that expected from the luminosity-metallicity relation alone.  All five galaxies in the pairs sample with extremely high central burst strengths have close companions and lower metallicities than the field galaxies.  This result provides strong circumstantial evidence that gas infall has plays an important role in the low metallicity galaxy pair members.
\item We observe an anti-correlation between nuclear metallicity and $\Delta(B-R)$.   The luminosity-metallicity residuals indicate that this anti-correlation is stronger than the relation expected from the luminosity-metallicity relation alone.
Galaxies which are actively growing bulges are likely to have bluer colors within their half-light radii compared to the colors of their outer disks.  Pairs with low metallicities are more likely to have blue bulges compared to higher metallicity galaxies; for metallicities below \OH$<8.8$, 33\% of galaxies have blue bulges while for metallicities \OH$>8.8$, 19\% have blue bulges.    In the blue bulge sample ($\Delta(B-R)>0$), a larger fraction of galaxies are in close pairs compared to the 
remainder of the sample (53\% c.f. 34\%).    This result provides additional evidence for gas flows in 
the low metallicity close pairs.
\end{enumerate}

We use the \OIIIOII\ vs \NIIOII\ diagram to separate ionization parameter from metallicity for our two 
samples and we show that the galaxy pairs have larger \OIIIOII\ ratios on average than the field 
galaxies.  This difference could result from either or a combination of (1) a larger ionization parameter from a larger fraction of young hot stars in the central regions of the galaxy pairs, or (2) the galaxy 
pairs experiencing a longer duration burst of star formation than the field galaxies. 

Finally, we show that the combination of metallicity gradients typical of late-type spiral galaxies 
and gas infall as predicted by merger simulations can reproduce the reduced nuclear metallicities that we observe in our close pairs.   In this scenario, galaxy interactions cause gas flows towards the central 
regions. The less enriched
gas from the outer regions dilutes the pre-existing nuclear gas to produce a lower metallicity than would be obtained prior to the interaction.   These gas flows trigger central bursts of star formation, producing strong central bursts, and possibly aiding the formation of blue bulges.  Observations of metallicity gradients and nuclear morphology in close pairs would test this scenario.

\acknowledgments
We thank the anonymous referee for constructive comments which helped to improve this paper.
We thank Josh Barnes, Dave Sanders, Lars Hernquist, Ken Freeman, Bob Joseph, and Fabio Bresolin for useful discussions.   We are grateful to Rolf Jansen for making the NFGS R-band magnitudes available to us.  L. J. Kewley is supported by a Hubble Fellowship.  M. J. Geller is supported by the Smithsonian Institution.  E. J. Barton acknowledges support from the University of California, Irvine.
This research has made use of NASA's Astrophysics Data System Bibliographic Services.

\appendix

\section{AGN contamination }\label{AGNcont}

In Section~\ref{class} we classify the NFGS and galaxy pairs using their optical emission-line ratios.
The \citet{Kewley01a} classification scheme removes galaxies that have an AGN contribution greater than $\sim 20$\% to their optical line ratios.   The potential residual 20\% from an AGN could alter the line ratios and thus the metallicities.  However, the majority of our pairs sample (99\%), including all of the close pair outliers have metallicities determined using the \NIIOII\ (77\%) ratio or a combination of the \NIIHa\ and \NIISII\ (22\%) line ratio.  An
AGN strengthens the \NII\ and \SII\ lines \citep{Baldwin81,Veilleux87,Kewley01a} but has little effect on the \OII\ line \citep{Ferland86,Ho93a,Ho93b,Ho05}.   Similarly, an AGN contribution increases the \NII\ line relative to \Ha\  \citep{Baldwin81,Veilleux87,Groves04b}.  Because both the \NIIOII\ ratio and the \NIIHa\ ratio increase with metallicity \citep{Denicolo02,Kewley02a,Pettini04}, an AGN contribution will cause our metallicities to be overestimated.  

The remaining 1\% of our sample  have metallicities determined using the \R23\ ratio.  We use the \R23\ ratio when log(\NIIOII)$\lesssim -1.0$.  This limit corresponds to a metallicity of \OH$\lesssim 8.4$.   In this low metallicity regime, the \R23\ ratio also rises with increasing metallicity.  An AGN contribution increases \OIII\ relative to \OII\ or \Hb\ \citep{Baldwin81,Veilleux87,Hao05,Ho05}, thus increasing the \R23\ ratio and causing the the resulting metallicity to be overestimated.

Therefore, a systematic contribution from a low-level AGN to the close pairs sample would cause an artificial increase in our metallicities which would shift the luminosity-metallicity relation in the opposite direction to that which is observed.   We conclude that an AGN contribution can not cause the observed shift in L-Z arelations toward lower metallicities for the close pairs.

\section{Morphological and Spectroscopic effects }\label{Other_effect}

\epsscale{0.7}
\begin{figure}
\plotone{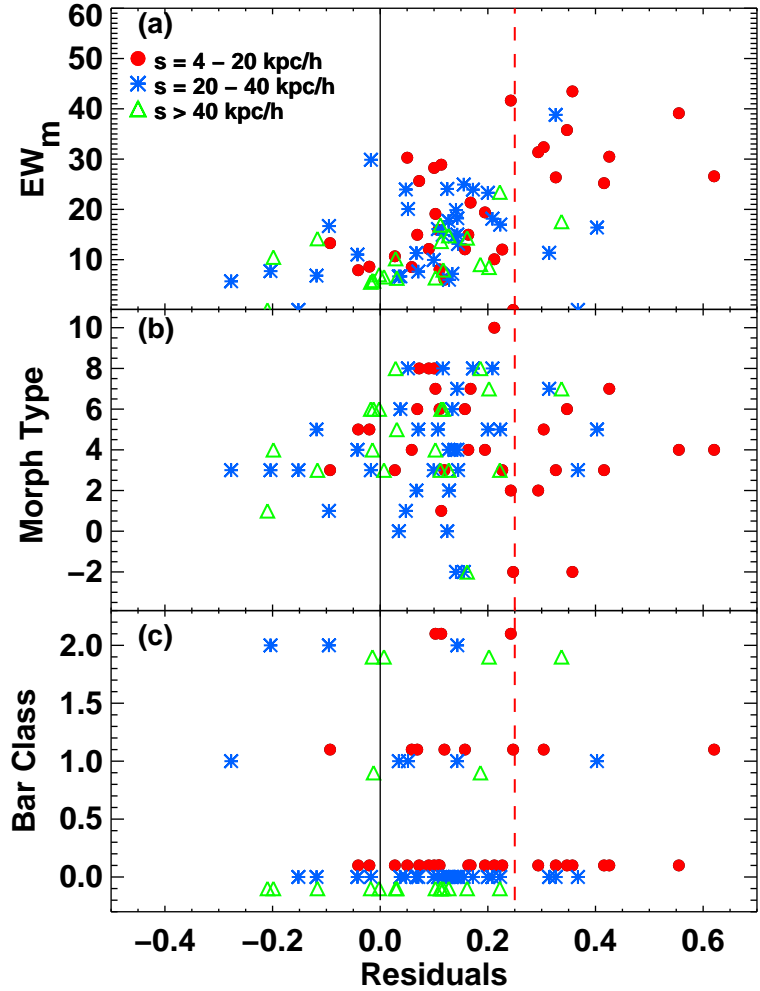}
\caption{The residuals of the galaxy pairs sample from the mean luminosity-metallicity relation for the 
field galaxies (Figure~\ref{LZplot}) versus (a) The mean equivalent width of the lines used to calculate metallicities for each galaxy, (b) de Vaucouleurs Morphological Type, (c) Bar Class.   Galaxies with strong bars, weak bars, and no bars are classed as 2,1,0 respectively.   We have shifted the close and widely separated pairs +0.1 and -0.1 from these bar classes respectively so that their distribution can be viewed more readily.   The black line shows where the pair members should lie if they were distributed 
uniformly about the least-squares fit to the field galaxies luminosity-metallicity relation.  The red dashed line shows the position of outliers in Figure~\ref{LZplot}.  Positive residuals correspond to lower-than-field metallicities.
\label{residuals}}
\end{figure}

In this section we investigate whether real morphological or spectroscopic effects might account for the shift in the luminosity-metallicity relation for close pairs.   
In Figure~\ref{residuals}a we plot the residuals 
of the galaxy pairs from the mean luminosity-metallicity relation of the field galaxies (dot-dashed line in Figure~\ref{LZplot})
 versus the mean equivalent width (EW$_{m}$) of the combination of lines used to determine metallicities.  For most galaxies in the sample, metallicities were determined using the \NIIOII\ ratio with 
reddening correction using the \Ha/\Hb\ ratio.  In this case, EW$_{m}$ is the mean equivalent width of  \NII, \Ha, \OII, and \Hb.  Galaxies with low EW$_{m}$ have the most uncertain metallicities.  The red dashed line marks the low metallicity outliers of the mean field galaxy LZ relation.  The EW${_m}$ for the 
outliers is high, indicating that the metallicities derived for them are not systematically affected by signal-to-noise issues.  

\citet{Kannappan04} analysed the morphologies of the pairs sample.
In Figure~\ref{residuals}b, we show the relationship between de Vaucouleurs morphological type and 
the residuals.  The outliers do not have significantly different range of types than the remainder of the 
sample.

Finally, we compare the presence of a bar with the residuals in Figure~\ref{residuals}c.  The presence of bars in the pairs sample was studied by \citep{Kannappan04}.  Galaxies with a strong bars, weak bars, and no bars are classed as 2,1,0 respectively.   The close and widely separated pairs are shifted 
+0.1 and -0.1 from these classes so that their distribution can be viewed more readily.   Figure~\ref{residuals}c shows that the fraction of bars in the outliers is low.  Thus, the presence of bars cannot account for the shift in the LZ relation for the close pairs.

\bibliography{library}

\end{document}